\documentclass{an}
\usepackage{graphicx}
\usepackage{times}
\Volume{00}              
\Year{0000}              
\Month{00}               
\Pagespan{000}{000}      

\begin{document}
\lhead[\thepage]{T. Miyaji et al.: XMM-Newton View of HDF-N and GWS Field}
\rhead[Astron. Nachr./AN~{\bf XXX} (200X) X]{\thepage}
\headnote{Astron. Nachr./AN {\bf 32X} (200X) X, XXX--XXX}

\title{XMM-Newton View of the Hubble Deep Field-North and Groth-Westphal
Strip Regions}

\author{T. Miyaji\inst{1} 
\and
R.E. Griffiths\inst{1}
\and
D. Lumb\inst{2} 
\and
V. Sarajedini\inst{3}
\and
H. Siddiqui\inst{2}
}

\institute{
Department of Physics, Carnegie Mellon University, 
5000 Forbes Ave., Pittsburgh, PA 15213, USA
\and
Science Payload Technology Divn., Research and Science Support Dept. of ESA
ESTEC,2200 AG Noordwijk, The Netherlands
\and 
Department of Astronomy, University of Florida, Gainesville, 
FL 32611-2055, USA
}
\date{Received {date will be inserted by the editor}; 
accepted {date will be inserted by the editor}} 

\abstract{We report the progress on our analysis of two deep
XMM observations in the areas of Hubble Deep Field-North and
the Groth-Westphal Strip field. We present the source detection method,
Monte-Carlo simulation on source detection, and our 
preliminary Log $N$ -- Log $S$ relations. Comparing these two fields
and other fields in literature, we find up to $\sim 30\%$ variations
in number counts near our flux limits, which is most likely to be due
to cosmic variance. This serves as a pilot study for a more systematic
investigation. The nature of the X-ray sources in the Groth-Westphal 
field is also discussed in terms of hardness ratios, morphology and
spectroscopic properties.  
}

\correspondence{miyaji@astro.phys.cmu.edu}

\maketitle

\section{Introduction}

With the large collecting area and field of view, XMM-Newton 
provides complementary probe to the faint X-ray source population
to the Chandra X-ray observatory. With a relatively short exposure,
it collects sufficient photons to provide decent X-ray spectroscopy.
Because of the limited fields of view in the Chandra Deep Fields
North and South, collecting intermediate exposure XMM-Newton
observations can probe the behavior of the X-ray sources
at several $\sim  10^{-16}$ [erg s$^{-1}$ cm$^{-2}$] (0.5-2 keV)
or several $\sim  10^{-15}$ [erg s$^{-1}$ cm$^{-2}$] (2-10 keV)
with fair sampling of the universe. 

 In this proceedings article, we report our progress on the analysis 
of two deep surveys using XMM-Newton, one on the Hubble Deep Field-North 
and surroundings (HDF-N) and the other on the Groth-Westphal Strip field 
(GWS). These fields have originally been selected because of the 
abundance of multiwavelength data. We note that the HDF-N field 
has the deepest Chandra observation (the Chandra Deep Field-North,
e.g. Alexander et al. this volume) and a part of the GWS has 200 ks of
Chandra observations. Because of the depth of these observations,
Chandra observations can provide better quality data than these 
XMM-Newton observations for most purposes. However, we use these
data as training sets for investigating source detection
properties with XMM-Newton and to extend the source count investigation
on multiple fields to investigate the effects of the large scale structure
of the universe. Also the XMM field of view of the GWS is much
larger than the single-pointing Chandra observation. 
Thus XMM-Newton still provides the only source of X-ray data for a 
significant part of the Groth Strip itself, consisting of 28 contiguous 
Hubble Space Telescope Wide Field Planetary Camera 2 (HST WFPC2) 
imaging observations, out of which 16 are within the XMM-Newton FOV.
 
\section{The XMM-Newton Observation}

The X-ray data presented here are from guaranteed time (GT) programs
of XMM-Newton. The observation log is shown below with the number 
of detected sources (within the central 11$^\prime$) and approximate 
flux limits.

\begin{table}[h]
\caption{The Log of Observations and Detected Sources}
\label{tab:log}
\begin{tabular}{cccccc}\hline
Field & \multicolumn{2}{c}{Expo. [ks]} & 
\multicolumn{3}{c}{$N_{\rm src} (S_{\rm lim}[{\rm erg\;s^{-1}\,cm^{-2}}]$)}\\ 
      &  PN & MOS & 0.5-2 keV & 2-10 keV & 5-10 keV\\
      
\hline
HDF-N & 155 & 180 & 158 & 116 & 49 \\
      & & & ($4\;10^{-16}$)&($2\;10^{-15}$)&($4\;10^{-15}$)\\
GWS   &  65 &  80 & 117 & 91 & 35\\
      & & & ($6\;10^{-16}$)&($4\;10^{-15}$)&($6\;10^{-15}$)\\ 
\hline
\end{tabular}
\end{table}

\section{Source Detection and Log $N$ -- Log $S$}

\begin{figure*}
\resizebox{\hsize}{!}
{\includegraphics[angle=270]{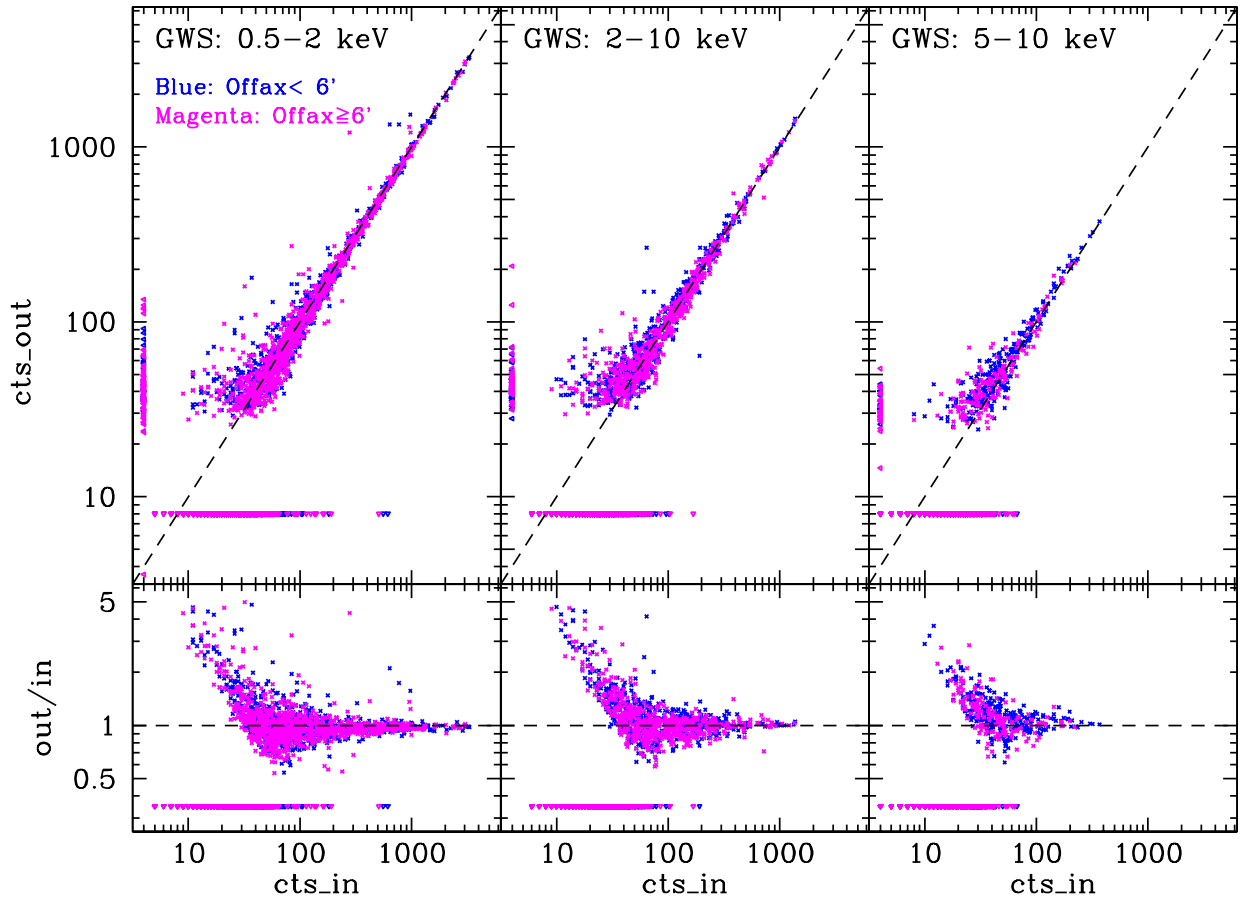}}
\caption{The source counts detected in the source detected chain
are plotted against the input counts for the Monte-Carlo simulations
corresponding to the XMM-Newton observations of GWS in three energy bands
as labeled. In the electronic version, the source within $<6\arcmin$ 
from the optical axis are plotted in blue points, while those outside 
in magenta. Also the ratio of output and input counts are shown.}
\label{fig:cts_cts}
\end{figure*}

\begin{figure*}
\resizebox{\hsize}{!}
{\includegraphics[angle=270]{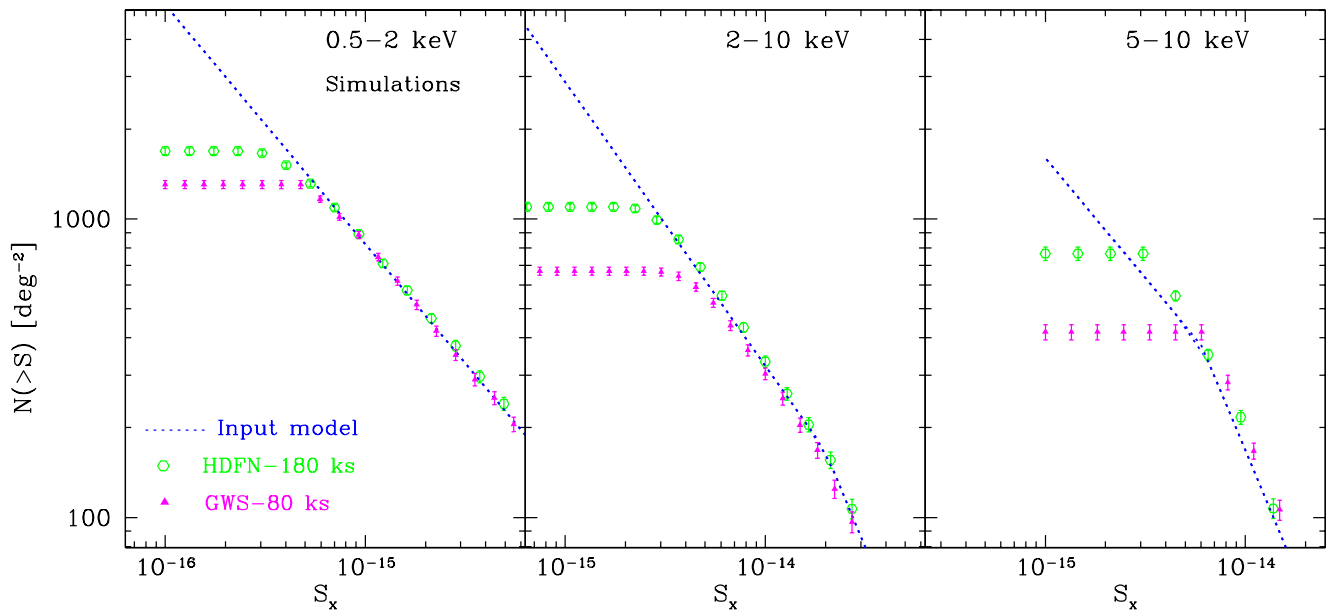}}
\caption{The input Log $N$ -- Log $S$ relation assumed for the Monte-Carlo
simulations (dotted lines) are compared with those derived from the source
detection procedure described in the text, assuming a uniform sensitivity
over the field }
\label{fig:sim_lns}
\end{figure*}

 We have performed source detection using the SAS procedure
to the sum of the PN and two MOS detectors. The exposure map
has been summed by weighting using PN and MOS responses in each
band (assuming a $\Gamma=1.4$ power-law). We have first used
{\em eboxdetect} in the local detect mode. We then created 
a smooothed background map using the {\em esplinemap}, which
gives a spline fitted image in the source excluded map. This 
is follwed by another {\em eboxdetect} run in the map detect
mode, where the sliding box cell searches for excesses over the
smoothed background map. Finally, the task {\em emldetect} has 
been applied to make maximum-likelihood fitting to the candidate
source positions. We have treated sources with the likelihood
value of $ML=-ln(1-P)>14$, where P is the probability that 
te source exists. 

There is a caveat in this source detection procedure. Depending on 
the selection of parameters such as the number of spline nodes, 
{\em esplinemap} often gives spurious wavy features, which give rise 
to a concentration of spurious sources or a void of detected sources. 
Currently we manually search for {\em espline} parameters which gives 
best results. This will be improved in the future.
 
 The inner 9.6 arcminute region has been used for the  Log $N$ -- Log $S$
analysis. In order to assess the limiting flux-area relation and 
incompleteness, we have made Monte-Carlo simulation using our multi-purpose 
image simulator {\em qimsim} (see Miyaji \& Griffiths 2002a). The simulated 
images have been processed through the same source detection procedure. 
Fig.\ref{fig:cts_cts} compares input and detected photon counts for the
simulation runs corresponding to the GWS observations. At the brightest
end, the input and output counts match quite well, even though the 
point spread functions used in the simulations and those for the 
{\em emldetect} correspond to different calibrations. This shows that
the counts in the source detection is not very sensitive to the PSF
calibration. We expect that the spread about the cts\_out = cts\_in
line decreases as we use an improved method for the background 
map generation (e.g. Baldi et al. 2002).  
Fig.\ref{fig:sim_lns} shows
the input Log $N$ -- Log $S$ relations and those calculated from
detected sources assuming a limiting flux corresponding to constant
count over the field.  Fig.\ref{fig:sim_lns} shows incompleteness near the
flux limit. Eddington bias can be seen in the 5-10 keV band, where
the input Log $N$-Log $S$ is still steep at the limiting flux.

\begin{figure}
\resizebox{\hsize}{!}
{\includegraphics[]{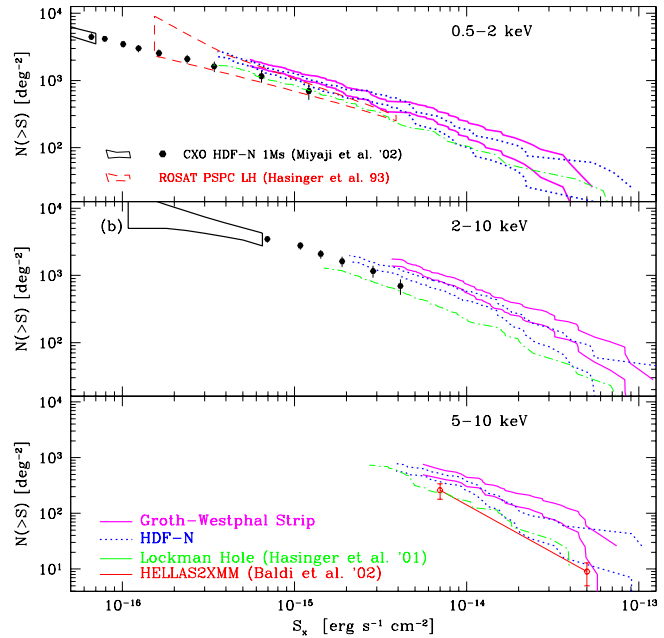}}
\caption{The Log $N$ -Log $S$ relations are plotted in 
the 0.5-2keV) (soft), 2-10 keV (hard) and 0.5-10 keV (ultrahard)
bands respectively for the XMM observations of HDF-North and
Groth-Westphal fields respectively. The upper and lower curves show
$\pm 1\sigma$ range. The results of a number of other sources are
also shown as labeled.}
\label{fig:lnls}
\end{figure}

 Fig.\ref{fig:lnls} shows the preliminary results, after first-order 
correction for incompleteness and Eddington bias using the results of the 
simulation. Further corrections similar to that of Moretti et al. 
(2002), which involves mapping of detected photon counts to incident 
counts for each source, will be applied in a future full-length paper. 
We also overplot the curves from XMM-Newton observation of the Lockman Hole 
(Hasinger et al. 1993) and 
of the inner part of the Chandra Deep Field North with fluctuation 
analysis results (Miyaji \& Griffiths 2002a). In the 5-10 keV plot,
the results from HELLAS2XMM is also shown. We observe the variation 
of 20-30\% at the limiting fluxes of our XMM-Newton observation, which
is most likely to be caused by the cosmic variance.   
   
\section{X-ray Sources in the Groth-Westphal Field}

In this section, we report the properties of the X-ray sources 
detected in the GWS field. We concentrate on the 24 sources within 
the HST WFPC2 fields of view, where the  DEEP\footnote{http://deep.ucolick.org/} redshift survey is also concentrated. A more extensive redshift survey 
(DEEP2 \footnote{http://astron.berkeley.edu/davis/deep/}) has started, which 
will cover most of the XMM-Newton field of view. This field is also 
a subject of the SIRTF legacy program, a planned target of the GALEX 
mission, and many other existing and future programs, which leaves potential
for unique studies of the X-ray source population.

\begin{figure}
\resizebox{\hsize}{!}
{\includegraphics[]{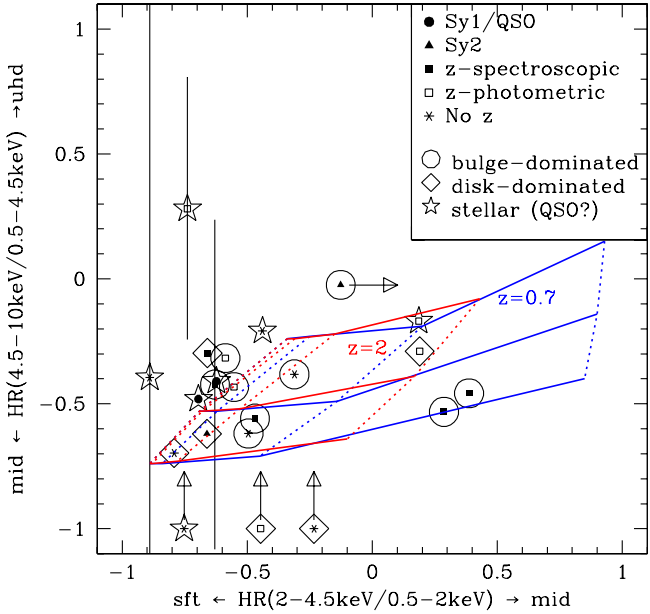}}
\caption{The X-ray color-color diagram of the 24 X-ray sources. The inner
symbol show spectral classification and the outer symbol show the 
morphological classification as labeled. Typical 1$\sigma$ errors 
are 0.05-0.2 and error bars are shown for those larger than 0.4. The grid
shows the location for absorbed power-law spectra for z=0.7 (wider grid,
shown in blue for the electronic version) and z=2 (narrower grid, shown in 
red for the electronic version). The solid lines correspond to photon
indeces of $\Gamma=$ 1,2, and 3 from top to bottom and the dotted lines 
to $Log N_{\rm H}=-\infty,21.5,22.5,$ and 23.5 cm$^{-2}$ from left to right 
respectively.}
\label{fig:hr}
\end{figure}

 Out of the 24 XMM-Newton sources, 10 have spectroscopic redshifts
from the DEEP survey and 5 have photometric redshifts (Brunner
et al. 2000). Out of the 10 spectroscopically identified sources,
two clearly have broad lines indicative of type 1 AGN, two have 
no broad permitted lines but show high excitation lines indicative
of Seyfert 2 activity. The rest have narrow emission lines with no
clear AGN indication. Using the HST Medium Deep Survey (MDS) database 
\footnote{http://archive.stsci.edu/mds/}, giving bulge/disk decomposition, 
we can immediately find the morphological properties of these sources as
shown in Table \ref{tab:morph}. Fig. \ref{fig:hr} shows X-ray color-color
diagram of these 24 sources with spectroscopic and morphological properties 
overlaid.   

\begin{table}
\caption{MDS Morphologies of X-ray sources }
\label{tab:morph}
\begin{center} 
\begin{tabular}{ccccc}
\hline
Stellar & Disk+Bulge & Pure Bulge & Galaxy$^{\rm a)}$ & No MDS \\
   3    &      2     &     3      &     2         &      2     \\
\hline
\end{tabular}
\end{center}

$^{\rm a)}$ An extended component has been resolved but the galaxy is 
  too faint for a disk/bulge discrimination.
\end{table}

In spite of the small number, we see some trends. Sources are distributed
along the grid, showing that absorbed power-law is a good description
of most sources. This can be seen with direct spectra of these sources 
(Miyaji \& Griffiths 2002b). We also note two bulge-dominated (elliptical)
sources with the hardest HR1, both of which are located at $z\sim 0.7-0.9$,
and with no apparent sign of AGN activity. In order to search for signs
of AGN activity, we plan to observe $\sim 10$ galaxies at $z>0.7$ with
Subaru OHS infrared spectrometer aiming at redshifted H$_\alpha$ and
[NII] emission lines. The propsal has been accepted and the observation
is expected to be in early 2003. We aim to find hidden AGN activity 
in the X-ray sources which only show normal galaxy characters in their optical
spectra.
  
\section{Conclusion}
 
 In this proceedings paper, we have reported the progress of our analysis on
deep XMM-Newton observations of the HDF-N and GWS fields. We have described
the source detection procedures, Monte-Carlo simulations, and resulting 
Log $N$ - Log $S$ relations in 0.5-2, 2-10, and 5-10 keV bands. Comparing
the results from these two fields as well as other works, we see 
the variations of source density of $\sim 30\%$ near our flux limit, which is most likely to be due to cosmic variance. With XMM-Newton, a more systematic study is possible by collecting archival data of the fields with similar depths.

 Currently our analysis on the GWS field is concentrated on the 
overlap with the WFPC2 strip, where the DEEP project has been concentrated. 
With the abundance of morphological data from the HST MDS, we can add insights into the nature of the X-ray sources. With our planned near-infrared spectroscopy using the OHS spectrograph attached to the Subaru 8.2m telescope, we expect to find indications of an AGN activity in X-ray sources which look
normal galaxies in the current optical data. We also can extend our spectroscopic identification with the DEEP2 program, which will cover most of the XMM-Newton FOV of GWS.

\acknowledgements

The authors acknowledge NASA grants NAG 5-10875 (LTSA) to TM 
and NAG 5-7423 to REG.

\end{document}